\documentclass[12pt]{iopart}

\usepackage{amssymb}
\usepackage{graphicx}
\usepackage{epsfig}

\begin{document}

\title{Quantum information transfer for qutrits}

\author{A. Delgado, C. Saavedra}
\address{Center for Quantum optics and Quantum Information, Departamento de F\'{\i}sica, Universidad de Concepci\'{o}n,
Casilla 160-C, Concepci\'{o}n, Chile.} \ead{aldelgado@udec.cl,
csaaved@udec.cl}

\author{J. C. Retamal}
\address{Departamento de F\'{\i}sica, Universidad de Santiago de Chile, Casilla 307, Correo 2, Santiago, Chile.}
\ead{jretamal@usach.cl}

\date{\today}

\begin{abstract}
We propose a scheme for the transfer of quantum information among
distant qutrits. We apply this scheme to the distribution of
entanglement among distant nodes and to the generation of
multipartite antisymmetric states. We also discuss applications to
quantum secret sharing.
\end{abstract}

\pacs{3.67.Mn, 3.65.Ud, 3.67.-a}

\section{\protect\bigskip Introduction}

Transfer and distribution of quantum states play an important role
in quantum communications and quantum information processing.
Quantum teleportation \cite{teleportation}, dense coding
\cite{densecoding}, and quantum key sharing \cite{qksharing}
require the distribution of a maximally entangled state among the
involved parties. Distributed quantum computing needs the transfer
of arbitrary quantum states among distant nodes. In this context,
an ideal scheme for the transfer of quantum states and the
distribution of entanglement was proposed in Ref
\cite{Cirac97,Mabuchi02}. This scheme considers ions stored in
distant cavities as qubits and the transfer of quantum states
takes place via a unidirectional exchange of photons triggered by
the application of laser pulses on the ions. The extension of this
work to the case of perfect transmission in presence of noise
channels was also studied \cite{vanEnk}. An important experiment
in this sense has been recently reported, were the recording of
quantum states of light onto an atomic memory with high fidelity
has been achieved \cite{memory}. Using photons propagating in
optical fibres, as flying qubits, and alkaline atoms, as
interfaces, transfer of quantum states between photons has been
reported \cite{interface}.

Recently, results from Quantum information theory have been
extended to the case of discrete systems of higher dimensions. For
instance, it has been shown that cold trapped ions systems, which
originally were proposed for implementing a qubit quantum
computer, can be extended naturally to implement a qutrit based
quantum computer\cite{guzman}. Another proposal for quantum
computations using qutrits has recently  been presented in spin
molecule systems \cite{Hugh}. Qutris have also been proposed to
improve the security of quantum key distribution \cite{durt}.

In this context arises naturally the question about transfer and
distribution of higher dimensional quantum states. In this work we
address this question. We restrict ourselves to transfer and
distribution of qutrits. We consider ions stored in two distant
high finesse optical cavities. The transfer of quantum states is
achieved through the emission and absorption of polarized photons
by cavities containing ions. We apply this scheme to the
distribution of entangled states among distant cavities. In
particular we study the generation of symmetric and antisymmetric
states and connect the results to the problem of quantum secret
sharing.

This article is organized as follows: In section \ref{sec:ideal}
we discuss the ideal quantum state distribution protocol. In
section \ref{sec:model} we outline the physical model and state
the scheme for the transfer of qutrit states. In section
\ref{sec:distribution} we discuss the distribution of entangled
states. In section \ref{sec:application} we present a protocol for
the generation of symmetric and antisymmetric states. In section
\ref{sec:qss} we discuss an application of the previous result to
the problem of quantum secret sharing. In section
\ref{sec:summary} we summarize and conclude.

\section{Ideal quantum state transfer}
\label{sec:ideal}

In the following we will consider two distinguishable qutrits.
Each qutrit is encoded in the electronic levels of a single ion
stored in a high finesse cavity. Our aim is to study the
conditions to achieve a perfect transfer of an arbitrary quantum
state between both qutrits, when the transfer is implemented via
the emission of polarized photons from one of the cavities and the
subsequent absorption of these photons by the other cavity.

The total system under study consists of both qutrits, the
quantized modes of the electromagnetic fields in each cavity and
the modes of the electromagnetic field connecting the cavities.
The states of the qutrits are given by
\begin{equation}
|\psi\rangle_k=c_0|0\rangle_k+c_{1,l}|1l\rangle_k+c_{1,r}|1r\rangle_k
\end{equation}
where $l$ and $r$ stand for left and right and $|0\rangle_k,
|1l\rangle_k$ and $|1r\rangle_k$ are three electronic levels of
the first ($k=1$) or the second ($k=2$) ion.

The initial state of the total system is given by
\begin{equation}
|\psi
_{s}\rangle=|\psi\rangle_1|0,0\rangle_1|\{0\}\rangle_e|0,0\rangle_2|0\rangle_2,
\end{equation}
where $|0,0\rangle_l$ denotes the vacuum state of quantized modes
at first ($l=1$) or second ($l=2$) cavity corresponding to
orthogonal polarized modes; $|\{0\}\rangle_e$ represents the
vacuum state of the environment modes between cavities.

Under appropriate conditions it is possible to transfer the state
of the qutrit to the modes of the first cavity, that is
\begin{eqnarray}
|\psi _s\rangle & \rightarrow & |0\rangle_1 \left( c_{0}|
0,0\rangle_1+c_{1l}|1,0\rangle_1+c_{1r}|0,1\rangle_1\right)
\otimes |\{0\}\rangle_e |0,0\rangle_2 | 0\rangle_2.
\end{eqnarray}
This transformation corresponds to a quantum state swapping
between the first ion and the modes of the first cavity. This can
be implemented if the electronic transitions
$|1l\rangle_1\rightarrow |0\rangle_1$ and $|2r\rangle_1\rightarrow
|0\rangle_1 $ evolve according to an effective Jaynes-Cummings
Hamiltonian (with an effective interaction time $\pi =(g\Omega
/\Delta )t$, each transition coupled to its corresponding cavity
mode.

Now we assume that both polarized photons are emitted from the
first cavity to the vacuum modes and subsequently absorbed into
the second cavity. Thereafter, the state is
\begin{eqnarray}
|\psi_s\rangle & \rightarrow &|0\rangle_1|0,0\rangle_1
|\{0\}\rangle_e \otimes \left(
c_0|0,0\rangle_2+c_{1l}|1,0\rangle_2+c_{1r}|0,1\rangle_2\right)
|0\rangle_2.
\end{eqnarray}
Thereby, it has been implicitly assumed that the photons are
perfectly absorbed at the second cavity.

Finally, the states of the modes in the second cavity are
transferred into the electronic levels of the ion, as in the first
step of the process. Thus, the final state of the total system is
given by
\begin{equation}
|\psi _{s}\rangle\rightarrow
|0\rangle_1|0,0\rangle_1|\{0\}\rangle_e|0,0\rangle_2|\psi\rangle_2,
\end{equation}
obtaining in this way a state where the information has been
perfectly encoded into the second ion.

However, we know that the photons emitted from the first cavity
have a nonzero probability of been reflected at the second cavity.
Thus, in a complete simulation of the transfer of the qutrit
state, we need time dependent Rabi frequencies of classical
fields. This allows for Raman processes which are described by
effective Jaynes-Cummings interactions.

\section{Physical setup and scheme for qutrit's state transfers}
\label{sec:model}

\begin{figure}[t]
\begin{center}
\includegraphics[width=0.50\textwidth]{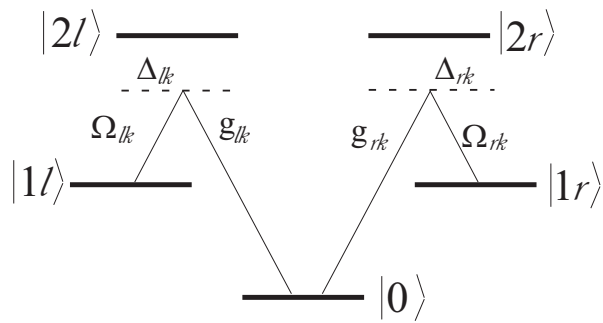}
\caption{Electronic level structure of trapped ion. Quantum
information of qutrits is stored in levels $\mid 0\rangle $, $\mid
1l\rangle $, and $\mid 1r\rangle $.} \label{fig:niveles}
\end{center}
\end{figure}

Qutrits can be physically realized using the electronic level
configuration currently found in experiments of lineal trapped
ions. For example, a suitable physical system are $^{138}$Ba ions,
where the qutrit can be defined in the metastable fine structure
corresponding to $4S_{1/2}$ and \ $5D_{3/2}$ levels. These states
can be addressed through Raman transitions such as the one shown
in figure \ref{fig:niveles}.

The Hamiltonian operator describing the unitary joint evolution of
the cavity modes and the five electronic levels of the ion is
\begin{eqnarray}
H_k & = & \hbar\sum_{j=l,r}\nu _{j,k}{a^\dagger}_{j,k} a_{j,k}
+\hbar \sum_{i=1}^5\omega_i|i\rangle_k\langle i|,
\nonumber\\
V_k & = & \hbar\sum_{j=l,r}
\left(g_{j,k}a_{j,k}|2j\rangle_k\langle 0|+
\Omega_{j,k}|2j\rangle_k\langle 1j|e^{-i\nu_{j,k}t}\right)+h.c.,
\end{eqnarray}
where the $k$ index denotes modes and levels at the first ($k=1$)
or second ($k=2$) cavity. Amplitudes $\Omega_{j,k}(t)=|\Omega
_{j,k}(t)|e^{-i\phi _{j,k}(t)}$ represent classical left ($j=l$)
and right ($j=r$) polarized pulsed laser fields connecting the
transition between levels $|1j\rangle_k$ and $|2j\rangle_k$. The
field operators $a_{j,k}$ and ${a^\dagger}_{j,k}$ describe the
transitions between levels $|1j\rangle_k$ and $|0\rangle_k$.

Assuming that laser and quantum modes are far detuned from the
corresponding optical atomic transitions, we can perform an
adiabatic elimination of the $|2j\rangle_k$  upper levels,
obtaining a pair of effective second order Hamiltonian operators

\begin{eqnarray}
H_{\textrm{eff}_{k}}=&-&\hbar\sum_j\Big(\delta
_{j,k}{a^{\dag}}_{j,k}a_{j,k}+ \frac{|g_{j,k}|^{2}}{\Delta
_{j,k}}{a^{\dag}}_{j,k}a_{j,k}|0\rangle_k\langle 0|
\nonumber\\
&+&\frac{|\Omega _{j,k}|^{2}}{\Delta _{j,k}} |j\rangle_k\langle
j|+i\frac{\Omega _{j,k}^{\ast }g_{j,k}}{\Delta _{j,k}}
a_{j,k}|j\rangle_k\langle 0|\Big) +h.c.,
\end{eqnarray}

where we have dropped out the label $1$ from the electronic levels
due to the fact that the $|2j\rangle_k$ states do not appear in
the above operator. These Hamiltonian operators allow us to
realize the necessary unitary operations to transfer an arbitrary
state of the qutrit to the quantum field modes.

Emission from the first cavity and subsequent absorption into the
second cavity can be described through cascaded open systems
formalism, where the Hamilton operator is given by
\begin{equation}
H_{nh}=-i\hbar\big(\sum_{j,k}\kappa_j \hat n_{j,k}+2\kappa_l
{a^\dagger}_{l,2} a_{l,1}+ 2\kappa_r {a^\dagger}_{r,2}
a_{r,1}\big),
\end{equation}
with $\kappa_{j}$ the decay rates of left ($l$) and right ($r$)
polarized modes of cavities $C_{1}$ and $C_{2}.$ The global
dynamics, including the information transfer into the first
cavity, photon emission from the first cavity, absorption in the
second cavity, and the final encoding in the second cavity can be
described by the Hamiltonian operator
\begin{equation}
H_{I}=H_{\textrm{eff}_{1}}+H_{\textrm{eff}_{2}}+H_{nh}.
\label{Htotal}
\end{equation}
Following reference \cite{mabuchi97}, we consider that no
reflected photons from the second cavity are to be detected.
Therefore, the following condition must hold:
\begin{equation}
(a_{j,1}+a_{j,2})| \Psi\rangle=0~~~j=l,r.
\end{equation}

In order to solve this problem, we consider only the Hilbert space
corresponding to the ions and the quantum fields during the
evolution. It is not difficult to realize that the vector state of
the global system can be written as
\begin{eqnarray}
|\psi _{s}\rangle &=&c_{0}a(t)|0\rangle_1|0\rangle_2
|0,0;0,0\rangle
\nonumber\\
&&+ c_{l}\left[ (b_{l,1}(t)|l\rangle_1|0\rangle_2 +
b_{l,2}(t)|0\rangle_1|l\rangle_2)|0,0;0,0\rangle\right.
\nonumber\\
&&+ \left. |0\rangle_1|0\rangle_2(d_{l,1}(t)|1,0;0,0\rangle +
d_{l,2}(t)|0,0;1,0\rangle)\right]
\nonumber\\
&&+ c_{r}\left[ (b_{r,1}(t)|r\rangle_1|0\rangle_2 +
b_{r,2}(t)|0\rangle_1|r\rangle_2)|0,0;0,0\rangle\right.
\nonumber\\
&&+ \left. |0\rangle_1|0\rangle_2(d_{r,1}(t)|0,1;0,0\rangle +
d_{r,2}(t)|0,0;0,1\rangle ) \right],
\nonumber\\
\label{fullstate}
\end{eqnarray}
where we have assumed the amplitudes defined such that
$b_{j,k}(t)=\alpha _{j,k}(t)e^{-i\phi _{j, k}(t)}$ with $j=l,r$
and $k=1,2$. In order to eliminate the dynamical Stark shifts, we
have assumed that the following conditions are satisfied
\begin{eqnarray}
  \delta_{j,k} &=& \frac{|g_{j,k}|^{2}}{\Delta_{j,k}}, \qquad  \dot{\phi}_{j,k}(t) = \frac{|\Omega _{j,k}|^{2}}{\Delta
  _{j,k}}
\end{eqnarray}
The state (\ref{fullstate}), together with the Hamilton operator
(\ref{Htotal}), leads us to the following system of coupled
equations for the amplitudes entering into Eq. (\ref{fullstate}):
\begin{eqnarray}
\dot{a}(t)&=&0,
\nonumber\\
\dot{\alpha}_{j,k}(t)&=&-\frac{\Omega _{j,k}g_{j,k}}{\Delta
_{j,k}} d_{j,k}(t),
\nonumber \\
\dot{d}_{j,k}(t)&=&+\frac{\Omega _{j,k}g_{j,k}}{\Delta
_{j,k}}\alpha _{j,k}(t)-\kappa _{j} (-1)^{k+1}d_{j,k}(t).
\label{setofequations}
\end{eqnarray}
The above equations have been simplified under the no detector
click condition for each polarization which ensures that
$d_{j,1}(t)+d_{j,2}(t)=0$. Thus, we have obtained two uncoupled
sets of equations, being each set associated with a particular
polarization. Each set turns out to be equivalent to the equations
obtained in Ref. \cite{mabuchi97} for the case of ideal
transmission of qubit states. To illustrate this let us write the
previous set of equations (\ref{setofequations}) with the
definitions $d_{j,1}=(d_{j,s}-d_{j,a})/\sqrt{2}$,
$d_{j,2}=(d_{j,s}+d_{j,a})/\sqrt{2}$ and the effective coupling
constant $\lambda _{l,k}=\Omega_{j,k}g_{j,k}/\Delta _{j,k}$. In
this case we obtain
\begin{eqnarray}
\dot{\alpha}_{j,k}(t)&=&(-1)^{k+1}\lambda
_{j,k}d_{j,a}(t)/\sqrt{2},
\nonumber\\
\dot{d}_{j,a}(t)&=&-\lambda_{j,1}\alpha_{j,1}(t)/\sqrt{2} +
\lambda_{j,2}\alpha_{j,2}(t)/\sqrt{2}. \label{set1}
\end{eqnarray}

In order to achieve an ideal transfer process we impose the
following conditions
\begin{eqnarray}
\alpha_{j,1}(-\infty ) &=&\alpha _{j,2}(\infty )=1~~~j=l,r
\nonumber\\
\phi _{j,1}(-\infty ) &=&\phi _{j,2}(\infty )=0~~~j=l,r.
\end{eqnarray}
Additionally, we must take into account the normalization
condition. Since the information transfer take place through
independent quantum channels, we assume independent normalization
conditions for the amplitudes associated with each channel, that
is
\begin{equation}
|\alpha _{j,1}|^{2}+|\alpha _{j,2}|^{2}+|d_{j,a}|^{2}=1~~~j=l,r.
\end{equation}

The above system of equations can be solved by imposing the
symmetric pulse condition \cite{mabuchi97} which, in the present
context, is given by $\lambda_{j,2}(t)=\lambda _{j,1}(-t)$. These
two conditions together with the set of equations (\ref{set1})
imply $\alpha_{j,1}(t)=\alpha _{j,2}(-t)$ and
$d_{j,a}(t)=d_{j,a}(-t)$. Considering the former relations and the
evolution equation $\dot{d}_{j,s}=0$, we obtain
\begin{equation}
\lambda _{j,1}(-t)=-\frac{\sqrt{2}\kappa _{j}d_{j,a}(t)+\lambda
_{j,1}(t)\alpha _{j,1}(t)}{\alpha_{j,2}(t)}\qquad t>0.
\end{equation}%
This expression allows us to determine the pulse shape by means of
the calculations in the first half of the time interval. This
solution can be in principle addressed once the initial conditions
are established. From the normalization condition, symmetry
considerations, and the equations $\dot{d}_{j,s}(t)=0$, the
following expression is obtained:
\begin{equation}
2\alpha _{j,1}^{2}(0)\frac{\lambda _{j,1}(0)^{2}+\kappa
_{j}^{2}}{\kappa _{j}^{2}}=1\qquad j=l,r.
\end{equation}%
Thus, the solution to the problem of transferring the state of a
qutrit follows from the transfer of the state of a qubit
\cite{mabuchi97}. This is a consequence of the information
channelling through independent dynamics of transitions
$\left\vert g_{k}\right\rangle _{a}$ $\rightarrow$ $\left\vert
e_{k1}\right\rangle _{a}$ and $\left\vert g\right\rangle _{a}$
$\rightarrow \left\vert e_{k2}\right\rangle _{a}$ at each cavity;
which turns out to be a consequence of the particular choice for
the interaction between the ions and the laser pulses.

\section{Distribution of entangled states of qutrits among distant nodes}
\label{sec:distribution}

The above scheme to transfer the state of a qutrit between distant
cavities can be applied to the distribution of entanglement among
distant qutrits. For sake of simplicity, we shall use the
following notation for the ion states $\mid 1l \rangle\rightarrow
\mid 1 \rangle$ and $\mid 1r \rangle\rightarrow \mid 2 \rangle$.

Let us consider for example the following state generated at
cavity $C_{1}$ between two ions $A$ and $B$:
\begin{equation}
\left\vert \Psi \right\rangle =\frac{1}{\sqrt{3}}\left( \left\vert
0\right\rangle _{a}\left\vert 2\right\rangle _{b}+\left\vert
1\right\rangle _{a}\left\vert 0\right\rangle _{b}+\left\vert
2\right\rangle _{a}\left\vert 1\right\rangle _{b}\right).
\end{equation}
A third ion C in its ground state is stored in a second distant
cavity $C_{2}$. If we apply the scheme for the transfer of qutrit
states between ions $B$ and $C$, we generate the state
\begin{equation}
\left\vert \Psi \right\rangle _{12}=\frac{1}{\sqrt{3}}\left(
\left\vert 0\right\rangle _{a}\left\vert 2\right\rangle
_{c}+\left\vert 1\right\rangle _{a}\left\vert 0\right\rangle
_{c}+\left\vert 2\right\rangle _{a}\left\vert 1\right\rangle
_{c}\right).
\end{equation}
Thereby, the entangled state between qutrits $A$ and $B$ has been
transferred to the ions $B$ and $C$.

We can also envisage the distribution of the entangled state of
three qutrits from one cavity $C_{1}$ to distant cavities $C_{2}$
and $C_{3}$. This process is carried out step by step distributing
one state at a time. In a physical implementation of this process
we should consider a two sided cavity $C_{1}$ in order to
distribute the states through each side of this cavity to cavities
$C_{2}$ and $C_{3}$. Thus, it is necessary to include the
possibility of controlling the transmission of the photons through
the mirrors in $C_{1}$.

\section{Generation of an Aharanov-Bohm state using qutrits}
\label{sec:application}

In the previous section we discussed the possibility of
distributing multiparticle entangled states among distant nodes.
In this section we study some special cases.

Let us start by studying the generation of symmetric and
antisymmetric states under permutations of individual particles.
The generations of such states can be implemented through unitary
local transformations and conditional operations. Here we choose
the Fourier transform $F^{(D)}$ and the control-not gate
$\textsc{XOR}^{(D)}$ for $D$-dimensional quantum systems
\cite{gxor}.

The conditional operation can be defined as
$\textsc{XOR}_{lmd}^{(D)}\left\vert i\right\rangle \left\vert
j\right\rangle =\left\vert i\right\rangle \left\vert i\ominus
j\right\rangle $ where $i\ominus j$ is the left modular difference
in $D$ dimensions. The operation defined in this way has the
property that the inverse operation is the same operation, that
is, $(\textsc{XOR}_{lmd}^{(D)})^{2}=I$. This is not true for an
alternative conditional operation defined through the modular
addition $\textsc{XOR}_{ma}^{(D)}\left\vert i\right\rangle
\left\vert j\right\rangle =\left\vert i\right\rangle \left\vert
i\oplus j\right\rangle ,$where the inverse is achieved as
$(\textsc{XOR}_{ma}^{(D)})^{D}=I$. An additional conditional
operation that can be defined for $D$-dimensional systems is the
right modular difference $\textsc{XOR}_{rmd}^{(D)}\left\vert
i\right\rangle \left\vert j\right\rangle =\left\vert
i\right\rangle \left\vert j\ominus i\right\rangle $; in this case
we have $(\textsc{XOR}_{rmd}^{(D)})^{D}=I$. These conditional
operations are related through local unitary operations.

In $D^2$-dimensional spaces the completely symmetric state can be
defined generalizing the idea of constructing the two particle
state given by:
\begin{equation}
\left\vert \Psi \right\rangle =\frac{1}{\sqrt{2}}(\left\vert
i\rangle _{1}|j\right\rangle _{2}+\left\vert j\rangle
_{1}|i\right\rangle _{2}).
\end{equation}%
There are two states which we can generate in two dimensions,
namely, the Bell states $\left\vert \Phi _{+}\right\rangle
=(\left\vert 00\right\rangle +\left\vert 11\right\rangle
)/\sqrt{2}$ and $\left\vert \Psi
_{+}\right\rangle =(\left\vert 01\right\rangle +\left\vert 10\right\rangle )/%
\sqrt{2}$. In both cases we must apply the operations
$\textsc{XOR}_{12}^{(2)}F_{1}^{(2)}\left\vert i\rangle
_{1}|j\right\rangle _{2}$ to the initial root states $\left\vert
00\right\rangle $ and $\left\vert 01\right\rangle $. This idea is
easily extended to $D$ dimensional spaces as follows:
\begin{equation}
\left\vert \Psi _{a_{1}a_{2}..a_{N}}\right\rangle
=U^{(D)}\left\vert a_{1}\right\rangle \prod_{j=2}^{N}\left\vert
a_{j}\right\rangle \label{syme1}
\end{equation}
\begin{equation}
U^{(D)}=\prod_{k=2}^{N}\textsc{XOR}_{1k}^{(D)}F_{1}^{(D)}.
\label{U1}
\end{equation}

In the case of two qubits we obtain completely symmetric states
under permutation of two particles. However, for $D$-dimensional
systems, the $U^{(D)}$ operation leads to states whose invariance
under permutations depends the initial states of the $N$
particles. In the case of qutrits the discrete Fourier transform
is defined as:
\begin{equation}
F_{n}^{(3)}\left\vert j\right\rangle _{n}=\frac{1}{\sqrt{3}}%
\sum_{l=0}^{2}e^{2i\pi lj/3}\left\vert l\right\rangle _{n}.
\label{fourier}
\end{equation}
The transformed states $\left\vert \bar{j}\right\rangle
=F_{n}^{(3)}\left\vert j\right\rangle $ read as
\begin{eqnarray}
\left\vert \bar{0}\right\rangle  &=&\frac{1}{\sqrt{3}}\left(
\left\vert 0\right\rangle +\left\vert 1\right\rangle +\left\vert
2\right\rangle \right)
,  \nonumber \\
\left\vert \bar{1}\right\rangle  &=&\frac{1}{\sqrt{3}}\left(
\left\vert 0\right\rangle +e^{2i\pi /3}\left\vert 1\right\rangle
+e^{-2i\pi
/3}\left\vert 2\right\rangle \right) ,  \nonumber \\
\left\vert \bar{2}\right\rangle  &=&\frac{1}{\sqrt{3}}\left(
\left\vert 0\right\rangle +e^{-2i\pi /3}\left\vert 1\right\rangle
+e^{2i\pi /3}\left\vert 2\right\rangle \right).
\end{eqnarray}
Let us consider the conditional $\textsc{XOR}_{lmd}^{D}\left\vert
i\right\rangle \left\vert j\right\rangle =\left\vert
i\right\rangle \left\vert i\ominus j\right\rangle $ operation.
Applying the $U^{(D)}$ transformation to initial registers
$\left\vert 0\right\rangle \left\vert 1\right\rangle \left\vert
2\right\rangle $ and $\left\vert 0\right\rangle \left\vert
2\right\rangle \left\vert 1\right\rangle $ we obtain the states:
\begin{eqnarray}
\left\vert \Psi _{012}\right\rangle  &=&\frac{1}{\sqrt{3}}\left(
\left\vert 0\right\rangle \left\vert 2\right\rangle \left\vert
1\right\rangle +\left\vert 1\right\rangle \left\vert
0\right\rangle \left\vert 2\right\rangle +\left\vert
2\right\rangle \left\vert 1\right\rangle
\left\vert 0\right\rangle \right)   \nonumber \\
\left\vert \Psi _{021}\right\rangle  &=&\frac{1}{\sqrt{3}}\left(
\left\vert 0\right\rangle \left\vert 1\right\rangle \left\vert
2\right\rangle +\left\vert 1\right\rangle \left\vert
2\right\rangle \left\vert 0\right\rangle +\left\vert
2\right\rangle \left\vert 0\right\rangle \left\vert 1\right\rangle
\right). \label{simepermu}
\end{eqnarray}

These states are invariant under cyclic permutations of the three
particle states of each individual system $(021\rightarrow
102\rightarrow 210$ and $012\rightarrow 120\rightarrow 201)$. They
are not invariant under the exchange of any two particle states.
More general states invariant under the exchange of any two
individual particles can be achieved by constructing
superpositions of the above states. This can be accomplished by
generating a Bell state starting from either the state $\left\vert
1\right\rangle \left\vert 2\right\rangle $ or the state
$\left\vert 2\right\rangle \left\vert 1\right\rangle $ $. $ In the
first case we generate the symmetric Bell state between particles
$2$ and $3$ so that
\begin{eqnarray}
\left\vert S\right\rangle
_{012} &=& \textsc{XOR}_{12}^{(3)}\textsc{XOR}_{13}^{(3)}F_{n}^{(3)}\textsc{XOR}_{23}^{(2)}H_{2}^{(2)}\left%
\vert 0\right\rangle \left\vert 1\right\rangle \left\vert
2\right\rangle  \nonumber \\
\left\vert S\right\rangle _{012} &=&\frac{1}{\sqrt{6}}(\left\vert
0\right\rangle \left\vert 1\right\rangle \left\vert 2\right\rangle
+\left\vert 1\right\rangle \left\vert 2\right\rangle \left\vert
0\right\rangle +\left\vert 2\right\rangle \left\vert
0\right\rangle
\left\vert 1\right\rangle   \nonumber \\
&&+\left\vert 0\right\rangle \left\vert 2\right\rangle \left\vert
1\right\rangle +\left\vert 1\right\rangle \left\vert
0\right\rangle \left\vert 2\right\rangle +\left\vert
2\right\rangle \left\vert 1\right\rangle \left\vert 0\right\rangle
).
\end{eqnarray}
Thereby, the resulting state is a superposition of $\left\vert
\Psi _{012}\right\rangle $ and $\left\vert \Psi
_{021}\right\rangle $, leading to the completely symmetric state
of three particles.

In the second case we generate the antisymmetric Bell state
between particles $2$ and $3$ so that
\begin{eqnarray}
\left\vert A\right\rangle
_{021} &=& \textsc{XOR}_{12}^{(3)}\textsc{XOR}_{13}^{(3)}F_{n}^{(3)}\textsc{XOR}_{23}^{(2)}H_{2}^{(2)}\left%
\vert 0\right\rangle \left\vert 2\right\rangle \left\vert
1\right\rangle \nonumber \\
\left\vert A\right\rangle _{021} &=&\frac{1}{\sqrt{6}}(\left\vert
0\right\rangle \left\vert 1\right\rangle \left\vert 2\right\rangle
+\left\vert 1\right\rangle \left\vert 2\right\rangle \left\vert
0\right\rangle +\left\vert 2\right\rangle \left\vert
0\right\rangle
\left\vert 1\right\rangle   \nonumber \\
&&-\left\vert 0\right\rangle \left\vert 2\right\rangle \left\vert
1\right\rangle -\left\vert 1\right\rangle \left\vert
0\right\rangle \left\vert 2\right\rangle -\left\vert
2\right\rangle \left\vert 1\right\rangle \left\vert 0\right\rangle
).
\end{eqnarray}
Thus we obtain a superposition of $\left\vert \Psi
_{012}\right\rangle $ and $\left\vert \Psi _{021}\right\rangle$,
leading to a completely antisymmetric state of three qutrits.

\section{Quantum state sharing with qutrits}
\label{sec:qss}

An interesting application of the states discussed in the previous
section arises in the context of quantum state sharing. The main
goal is the splitting of a quantum state among several parties
such that a subset of the parties can recover the original state
only if all the parties agree to cooperate. The state to be shared
can be considered as the key to active some process and the scheme
attempts to control the misuse of the key by potentially dishonest
parties. A quantum state sharing protocol can be formulated using
the state $\left\vert \Psi \right\rangle _{021}$, whose generation
has been studied in the above section.

The state to be share is
\begin{equation}
\left\vert \chi \right\rangle _{3}=c_{0}\left\vert
0\right\rangle_3 +c_{1}\left\vert 1\right\rangle_3
+c_{2}\left\vert 2\right\rangle_3.
\end{equation}
The joint state of the four qutrits is $\left\vert \chi
\right\rangle _{3}\left\vert \Psi \right\rangle _{021}$. Qutrit
zero and three belongs to the first party, qutrit two belongs to
the second party and qutrit three to the third party. Each party
corresponds to a cavity with the qutrits physically implemented as
trapped ions. The joint initial state obeys the following
identity:
\begin{eqnarray}
F_1^{(3)}|\chi\rangle_3|\Psi\rangle_{021}=&&\frac{1}{3\sqrt{3}}\sum_{m,\mu}|\Psi_{m,\mu}\rangle_{30}
\sum_l Z_1^{1-\mu}|l\rangle_1
\nonumber\\
&& \otimes X_2^{2-\mu}Z_2^{m+l}|\chi\rangle_2. \label{identity}
\end{eqnarray}
The states $|\Psi_{m,\mu}\rangle_{03}$ denote the states of
qutrits zero and three defined by
\begin{equation}
|\Phi _{m,\mu}\rangle_{30}
=\textsc{XOR}_{30}^{3}F_{3}^{3}\left\vert m\right\rangle
_{3}\left\vert \mu\right\rangle _{0}.
\end{equation}
These states form a set of mutually orthogonal, maximally
entangled states and can be considered as the generalization of
the Bell basis to two qutrits. The operators $X$ and $Z$ are
defined as
\begin{equation}
X=\sum\limits_{n=0}^{2}|n+1\rangle\langle n|, \quad
Z=\sum\limits_{n=0}^{2}\omega(n)|n\rangle\langle n|,
\end{equation}
where $\omega(n)$ are roots of the unity, i.e, $\omega \left(
n\right) =\exp \left( \frac{2\pi i}{D}n\right)$.

The transference of the quantum state of qutrit zero to qutrit two
can be read from Eq. (\ref{identity}). A generalized local Bell
measurement on qutrits zero and three projects qutrits two and one
to the state
\begin{equation}
\sum_l Z_1^{1-\mu}|l\rangle_1 X_2^{2-\mu}Z_2^{m+l}|\chi\rangle_2.
\end{equation}
Applying a Fourier transform to qutrit one and measuring it, the
state $\left\vert \chi \right\rangle$ is transferred to qutrit
three modulo a unitary transformation, that is
\begin{equation}
X_2^{2-\mu}Z_2^{m+l}|\chi\rangle_2,
\end{equation}
which depends on the outcome $(m,\mu)$ of the Bell measurement and
on the result $(l)$ of a measurement on qutrit one. In order the
third party can recover the state $|\chi\rangle$ on his qutrit,
the other parties must agree to share the measurement results.

\section{Summary}
\label{sec:summary}

In this work we have studied the problem of information transfer
between quantum systems embodying information in three-dimensional
Hilbert spaces. As we have seen, there is an ideal protocol for
information transfer having a possible  experimental system which
could be used to implement it. This physical realization is an
extension to qutrit systems of the information transfer proposal
between qubits embedded in high finesse optical cavities. As we
have shown in a qutrit system, the physical implementation works
as if we were transferring information between two independent
qubit channels. In addition, we have discussed the possibility of
generating multipartite states among these three-dimensional
quantum systems and we have discussed the possibility of using
them for quantum state sharing.

\section{Acknowledgement}

This work was supported by FONDECYT Grants Nos. 1030189 and
1040591 and by Milenio ICM P02-49F.

\end{document}